**A National Research Agenda for Intelligent Infrastructure: 2021 Update**

Daniel Lopresti             Shashi Shekhar
Lehigh University      University of Minnesota

**Executive Summary**

Strategic, sustained Federal investments in intelligent infrastructure will increase safety and resilience, improve efficiencies and civic services, and broaden employment opportunities and job growth nationwide. The technologies that comprise intelligent infrastructure can also provide keys to solving some of the most vexing challenges we face today, including confronting future pandemics and natural disasters, achieving sustainability and energy efficiency goals, and advancing social justice. Enabling those technologies effectively will require investment in the associated computing research as well, beyond and in concert with the basic building projects. In 2017, the Computing Community Consortium (CCC) produced a series of intelligent infrastructure whitepapers, and in 2020 CCC issued a set of companion whitepapers on closely related topics. Here we briefly survey those earlier works, and then highlight four themes of rising  national prominence where intelligent infrastructure can also play an enabling role, driven by experiences with the COVID-19 pandemic and the social justice movement. We conclude with recommendations for the necessary research investments.

**The Research Landscape in Intelligent Infrastructure**

Intelligent infrastructure is the embedding of sensing, computing, and communications capabilities into traditional urban and rural physical infrastructures such as roads, buildings, and bridges, along with a wide range of techniques to assemble and analyze the data that is generated and make it useful to decision makers and society as a whole. In 2017, the Computing Community Consortium (CCC) produced a series of 10 whitepapers that made the case for significant Federal investment in intelligent infrastructure, identifying a number of major challenges that would have to be addressed including basic research, support for community/industry partnerships, and the special role played by universities in charting the path forward. These earlier whitepapers focused on the necessary computing research in next-generation transportation [B17]; the smart grid [C17]; disaster management, public safety, and community resilience [D17]; smart cities [E17]; smart agriculture [F17]; cybersecurity [G17] and privacy [J17]; intelligent infrastructure in a rural context [H17], and wireless networking [K17].

An overarching whitepaper [A17] argued that building for the 21st Century and beyond demands the incorporation of intelligence at the very foundations of our infrastructure. To do otherwise would deprive the U.S. of the significant advancements this would bring to society, including increased safety and resilience, improved efficiencies and civic services, and broader employment opportunities and job growth. Infrastructure is the springboard for international economic competitiveness, and a race in which we cannot afford to fall behind. As noted in that earlier whitepaper, sustained Federal investment in the nation's infrastructure after World War II led to an era of increasing prosperity for the country. We are now on the cusp of a similar opportunity, and intelligent infrastructure must play a central role.

Much of what was written in the CCC's 2017 whitepapers remains relevant today. Here we highlight four key themes that have recently risen to national attention as a result of the tumultuous events of 2020: the COVID-19 pandemic and worldwide natural disasters, sustainability and energy efficiency, job recovery and employment opportunities, and the advancement of social justice. For each of these existential challenges to our society, intelligent infrastructure can be seen as a key enabler for the solutions we seek. After this overview, we provide a list of recommendations for investments in the associated computing research. At the end of this whitepaper we provide a short bibliography of CCC's 2017 intelligent infrastructure whitepapers, along with a companion set of whitepapers released in November 2020 in the areas of AI and machine learning, privacy and security, robotics, networking, and informatics; technologies that will support intelligent infrastructure and where research investments are also required.

**The Nexus Between Intelligent Infrastructure and Social Justice**

Recent events have highlighted a set of important interrelated challenges: the growing presence of public surveillance by governmental and commercial entities and the attendant privacy issues, along with the rising demands for social justice. The latter can either be advanced or diminished by new-found capabilities to monitor the actions of both public servants and the public, and hold those who violate the law accountable. Body-worn video cameras for law enforcement provides an early example of this, but widespread deployment of intelligent infrastructure will be a key enabler in the transition and provides the classic "double-edged sword": safety, security, and civil rights versus the privacy of the individual. These ethics and technology issues are best approached from an interdisciplinary perspective [C20].

AI analytics at the edge of pervasive networks in smart and connected communities could improve privacy by maintaining a level of anonymity when implemented properly, but more research is still needed [A20, R20]. Experiences with the abuse of private data on the commercial side hints at what we can expect when ubiquitous government surveillance in physical spaces becomes prevalent [J20]; it is not clear who will own the data captured while watching you drive down a public street: the county government? The vendor who installed and operates the system monitoring traffic on the roadway? The company storing the data and providing the analytics? How can effective policies be developed and adherence guaranteed? These are questions that also require research.

Other applications may prove less controversial. The embedding of pervasive environmental sensors across a city landscape, including in underprivileged sections of the city, and making the data public, will allow the identification, monitoring, and remediation of air quality issues that have frequently plagued the less fortunate in our society [E17], again assuming key research issues can be addressed.

Another pressing issue, even before the pandemic but certainly exacerbated by it, are the disparities in broadband access across the U.S., which presents significant barriers to achieving equality in employment, education, and healthcare. This is especially an issue in rural and tribal areas [H17], but even some cities still have problems with equal access to broadband for all of their residents. Advances in next-generation wireless networking have the potential to help [R20], but much research is also needed in developing new and better ways to deliver vital services to remote users (e.g., [K17]). A related realization and topic for ongoing research is the recognition that many of the accommodations developed to support people with disabilities in the physical world (often mandated by Federal law) do not automatically carry over to the online world; more research needs to be done here as well. Finally, the recognition that underserved communities are often the hardest hit when disasters strike provides another opportunity where intelligent infrastructure can be applied to help level the playing field and create a more equitable, supportive living environment [D17].

**Resilience to Pandemics and Other Disasters**

The COVID-19 pandemic, the recent devastating wildfires in the American West and elsewhere, and the terrorist attacks of 9-11 remind us that modern society is not immune from natural disasters (e.g., infectious disease outbreak, hurricanes earthquake) and man-made disasters (e.g., terrorism, social unrest), and their large-scale consequences (e.g., fiscal, social, built and natural environment). Computing and intelligent infrastructure are key to enhancing resilience by improving preparation and plans for responding to and recovering from and effectively adapting to high-risk disasters [K20]. Computing breakthroughs ranging from broadband to robotics to spatial computing (e.g., COVID-19 maps, contact notification apps, local and regional disease transmission-dynamics models) to GeoSocial media (e.g., geo-targeted amber alerts, Ushahidi, OpenStreetMap, tweets, facebook, tiktok) have provided game-changing capabilities and become key tools for disaster management [D17]. For example, the broadband telecommunication infrastructure proved invaluable to enabling stay-at-home orders during the COVID-19 pandemic by supporting tele-education, tele-health, and remote work. In addition, computational simulation models (e.g., SEIR COVID-19 spread models) were useful in projecting COVID-19 trajectories to protect hospital capacity. In addition, Wireless Emergency Alerts that are used to warn geo-targeted people at risk of flash floods, dust storms, hurricanes, etc. are now being used to inform people to take actions to reduce their COVID-19 exposure and risk [D17].

As a result many compelling and life changing computing opportunities have recently emerged. For example, there is a critical need to strengthen the *National Pandemic Informatics Infrastructure* [K20] by modernizing the public-health databases and information systems as well as provisioning of high performance computing resources to support simulation of disease transmission dynamics models and analysis of disease big data. Other compelling opportunities include next generation computational simulation leveraging AI [D20] and spatially-detailed *Digital Twins* (big-data and physical/social science driven computational models of our world and society), which can help preparation and planning by projecting the impact of disaster scenarios and comparing benefits from alternative interventions. The cyber vulnerabilities of grid components must also be characterized and mitigated to prevent cyber threats. Privacy of sensitive power system data should be preserved by efficient data protection and encryption techniques [C17]. Basic computing research is necessary to address many of these open issues.

## Addressing the Impact on Jobs and Employment

At the same time the move to intelligent infrastructure increases efficiencies and lowers costs for communities, improves public safety and security, and preserves natural resources and enhances sustainability, there will be a significant impact on jobs across a wide variety of sectors. Some of this impact will be negative, with whole job categories disappearing due to automation, while other aspects of the impact will be positive, with new careers arising that do not exist today, many in high tech fields. Addressing this impact to maintain and enhance employment opportunities will require a substantial coordinated effort involving government, industry, and academia, the latter both to address open research challenges as well as to provide the educational services necessary to "re-tool" workers to be successful in the greener careers of tomorrow made possible by intelligent infrastructure. We see tremendous potential for humans to work side-by-side with AI systems and robots, increasing productivity, job safety and satisfaction, and international competitiveness.

As noted in the 2017 version of this whitepaper [A17], future communities infused with intelligent infrastructure will look and act very different from today. Some changes will be dramatic and especially disruptive, such as the transition to autonomous vehicles, while others will flow more naturally and indeed have even started taking place, accelerated by current pandemic (highly automated grocery store check-outs, smartphone-based commerce-at-a-distance replacing physical interactions of the past, e.g., feeding a parking meter). As we observed in 2017, investments in lifelong learning and technical training will be critically important to help workers remain employed and take full advantage of the benefits provided by these new technologies. The Internet of Things (IoT) is sometimes considered synonymously with intelligent infrastructure, and as more and more AI is embedded in these so-called edge devices [A20], worker training will be needed here as well.

Even though the phrase "intelligent infrastructure" may conjure up images of the smart city of tomorrow, in our earlier whitepapers we have also discussed the tremendous positive impacts on agriculture and rural communities and the job opportunities there as well [F17, H17]. Once again, embedded intelligence brings on sustainability and productivity gains that will be a critical contribution to the international competitiveness of U.S. agriculture. And, likewise, there will be a significant demand for continuing education so that workers can be prepared for these major changes. In [H17] we also observed that the growth of the digital economy in an increasingly connected world would lead to new remote work opportunities that could benefit rural communities with their lower costs of living and higher quality of life. Indeed, the pandemic has accelerated this transition in dramatic fashion; the net benefit of remote work opportunities on depressed rural communities could be a notable positive at a time of so many negatives.

Robotics has the potential to magnify and amplify the capabilities and productivity of teams involving both humans and robots. As we discuss in our 2020 whitepaper [M20], this does not necessarily mean the introduction of robots will lead to fewer jobs. Indeed, by some measures employment in industries making greater use of robots has actually increased. We must also focus, however, on the nature and the quality of those jobs, and matching workers' skills with the increasing technical demands of maintaining and operating their robot "teammates," so that the work available to humans is rewarding and not menial. Research in human-robot teaming can help with this, as can investments in education.

Next-generation vehicles come in a variety of different types: automated, connected, eco/electric, and shared (ACES). The sustainability and productivity gains from these vehicles will be substantial as will the economic benefits, but attention must be paid both to the enabling research and also to helping the workers who will be displaced if they are not supported in their attempts to adapt to the new demands of the day.

Cooperation between AI and humans is a common theme across a range of intelligent infrastructure applications. Human workers must cooperate with each other to do their jobs, and if AI is to be a good teammate it must be able to cooperate with humans and other AIs, too. This will require AI that understands human preferences, can model ethical behavior, and operates within current laws and regulations. The AI must be trustworthy and helpful at all times, avoiding any form of manipulation that could mislead or even endanger human co-workers. All of these issues are topics of current research and have important implications for supporting humans in the workplace as well.

## Sustainability and Energy Efficiency

Our planet is facing challenges ranging from environmental degradation to climate change. Life threatening weather events (e.g., floods, droughts) are becoming more extreme and frequent. Consequently, the United Nations has identified a set of sustainable development goals including food, energy and water security, climate action, sustainable communities, biodiversity, and addressing inequality. There are a number of social justice aspects to consider with these as well. Poor communities are often located in the most risky areas in terms of climate disasters. They also need the most support when disaster strikes. Computing is crucial to addressing these challenges facing our changing planet [F17]. For example, sensor networks and new methods to analyze their data are needed to monitor our planet to help us understand biological, physical, and social changes. Computational simulations may help forecast rates of sea level change, loss of polar ice, and predict critical geo-events. It is also important for societal priorities including security, public health, smart cities, transportation, climate, environment, food, energy, water [A17]. Thus, Microsoft announced that the goal of being Carbon negative (removing more carbon than they emit) by 2030 and launched "AI for Earth" to address global environmental challenges. In addition, multiple planetary-scale platforms for Earth data have emerged including the Earth on Amazon Web Services, the Google Earth Engine, and the NASA Earth Explorer. This high-frequency and high spatial resolution data is being used to monitor global crops to avoid future Arab spring like events. Furthermore, researchers are becoming aware of the impacts of research activities, such as the greenhouse gases and energy use associated with cloud computing as well as training of very large AI models (e.g., GPT-3), and a conference-driven publication culture [E20]. While industry involvement is a valuable piece of the puzzle, the broad importance of these issues to our society calls out for Federal investments as well in independent, unbiased academic research. For example, it can catalyze a drive to become carbon-negative, if not carbon negative. Further, a global-dashboard to monitor the health of our planet may support early detection of unsustainable events (e.g., deforestation) and trends (e.g., shrinking Aral sea) towards timely interventions.

It should be noted that computing to address climate and environmental challenges, namely, spatial computing, has unique characteristics [F17]. For example, there are significant interactions among the vexing geo-challenges. Optimizing solutions for one problem may have unintended side-effects for other problems, such as the increased use of fertilizers for food production have adversely affected water quality. Current computational models need to be viewed as approximations to geo-phenomena, whether physical, biological, or cultural. Progress is made by either improving the approximation or reducing computational cost or both. In addition, the training of machine learning and geospatial AI models faces great challenges due to non-stationarity of Earth phenomena (e.g., climate change) as well a lack of training samples as significant geo-events can be very rare and ground truth data are labor-intensive and time-consuming to collect. Thus, it is timely for a new computing research initiative to engage the broad computing community along with complementary disciplines to better understand and protect our changing planet. Among the risks is the substantial carbon footprint required for developing current AI technologies, including large machine learning models. Intelligent infrastructure opens the door to dramatically better energy efficiency and sustainability [A20, D20].

**Call to Action**

Infrastructure is vital to our nation's ongoing health, wealth, and well-being. The time is ripe for substantial, sustained Federal investments in basic infrastructure (e.g., transportation, power grid, utilities, networking) for both maintenance and modernization to maintain our global economic competitiveness. Such investments will also provide a much-needed economic stimulus to accelerate the post-COVID recovery by creating new jobs in traditional engineering and construction fields, supporting associated short-term and long-term economic growth.

However, infrastructure has a long lifespan and thus it is very important for infrastructure investments to be made with a keen eye toward the future. For illustration, Table 1 lists transportation trends along with some associated infrastructure recommendations. First, vehicles are increasingly automated, connected, environment-conscious (e.g., electric) and shared. This demands a move toward an Internet of Infrastructure Things (IoIT) approach, with increased sensing, communication, computing and data analytics (e.g., AI) capabilities. Second, transportation ranks among the largest sources of greenhouse gas emissions which drives climate change. This status quo is not healthy or acceptable, and future transportation infrastructure should reduce its carbon footprint by using sustainable technologies. For example, street navigation apps may be encouraged to recommend routes that reduce emissions and energy use, which will further the appeal and adoption of electric vehicles. Third, COVID-19 demonstrates the need for infrastructure resilience as our society has realized its massive dependence on broadband for tele-work, tele-health and tele-education. It has also highlighted significant inequalities across neighborhoods and communities, including rural and tribal areas that have limited access to broadband. The pandemic has also highlighted the importance of long-haul transportation (e.g., trains, trucks) as well as last mile on-demand delivery (e.g., doordash, UberEats) during stay-at-home orders as a valuable measure in controlling disease spread. Other disasters during 2020, including widespread flooding due to hurricanes and destructive wildfires, also argue in favor of building infrastructure that is resilient; intelligence adds this. Furthermore, low-income neighborhoods often have weaker transportation infrastructures leading to long travel times to access jobs, healthcare, education, and other basic services. In addition, there is often a geographic mismatch between jobs and affordable housing. Future infrastructure should work to reduce existing inequalities and provide strong evidence to allow us to monitor progress. Fifth, there is a growing misalignment between the jobs of the future and our current workforce. We need to create education and training programs to provide the needed skills and knowledge (e.g., in areas relating to the design, deployment, and uses of intelligent infrastructure, AI, and cyber-security) so that American workers are prepared for the future.

*Table 1: Example Trends and Infrastructure Recommendations*

| | Trends | Intelligent Infrastructure recommendation |
|---|---|---|
| 1. | Automated, Connected, Environment-conscious and Shared (ACES) Vehicles | Move towards an Internet of Infrastructure Things (IoIT) by increasing sensing, telecommunication (e.g., broadband), computing (cloud, edge) and data analytics including AI on highways and other critical infrastructure |
| 2. | Sustainability, Energy Efficiency: Transportation is a major source of greenhouse gases and driver of climate change | Reduce (transportation) infrastructure carbon footprint via sustainable technologies such as electric vehicles and eco-routing service which recommend routes and schedules to reduce energy use and emissions. |
| 3. | Increasing frequency and magnitude of natural and man-made disasters | Universal access to broadband for tele-commuting, tele-health, distance education, etc. to improve resilience to natural and man-made disasters |
| 4. | Increasing inequality, needs for social justice | Intelligent infrastructure to improve quality of life and reduce inequalities in access to jobs, healthcare, education and other essential activities |

| 5. | Structural mismatch between jobs and skills in labor market | Grow education and training in intelligent infrastructure to create workforce of the future |
|---|---|---|

**Importance of Federal Leadership and Investment**

The Federal Government has a crucial role in shaping intelligent infrastructure investments and deciding the foundations on which the national economy will be built going forward. As we have noted above (and in our previous whitepapers), there is a significant amount of basic research required to achieve the promise of intelligent infrastructure. Some of this research can be resourced through current programs such as NSF's Smart and Connected Communities in conjunction with the MetroLab Network spanning numerous cities, counties, local governments and universities. However, existing programs provide only modest investments in basic research and are not of a sufficient scale to support the broad, national transition we must make to achieve the full benefits of intelligent infrastructure. Any such effort must leverage our world-class universities via Federal funding of basic research and development to invent, deploy, and evaluate the necessary new technologies, to participate in the all-important discussions of policies and ethics, and to train the intelligent infrastructure workforce of the future.

## November 2020 Whitepapers

| Artificial Intelligence | |
|---|---|
| A20 | *Artificial Intelligence at the Edge* |
| B20 | *Artificial Intelligence and Cooperation* |
| C20 | *Interdisciplinary Approaches to Understanding Artificial Intelligence's Impact on Society* |
| D20 | *The Rise of AI-Driven Simulators: Building a New Crystal Ball* |
| E20 | *Next Wave Artificial Intelligence: Robust, Explainable, Adaptable, Ethical, and Accountable* |
| **Socio-Technical Computing** | |
| F20 | *An Agenda for Disinformation Research* |
| J20 | *Modernizing Data Control: Making Personal Digital Data Mutually Beneficial for Citizens and Industry* |
| **Broad Computing** | |
| K20 | *Pandemic Informatics: Preparation, Robustness and Resilience* |
| L20 | *Infrastructure for Artificial Intelligence, Quantum and High Performance Computing* |
| M20 | *Robotics Enabling the Workforce* |
| N20 | *A Research Ecosystem for Secure Computing* |
| **Core Computer Science** | |
| P20 | *Post Quantum Cryptography: Readiness Challenges and the Approaching Storm* |
| Q20 | *Theoretical Computer Science: Foundations for an Algorithmic World* |
| R20 | *Computing Research Challenges in Next Generation Wireless Networking* |
| S20 | *Advancing Computing's Foundation of US Industry & Society* |

## March 2017 Whitepapers

| Intelligent Infrastructure | |
|---|---|
| A17 | *A National Research Agenda for Intelligent Infrastructure* |
| B17 | MOBILITY21: Strategic Investments for Transportation Infrastructure & Technology |
| C17 | Digital Grid: Transforming the Electric Power Grid into an Innovation Engine for the United States |
| D17 | Research Agenda in Intelligent Infrastructure to Enhance Disaster Management, Community Resilience and Public Safety |
| E17 | City Scale Intelligent Systems and Platforms |
| F17 | Intelligent Infrastructure for Smart Agriculture: An Integrated Food, Energy and Water System |
| G17 | Safety and Security for Intelligent Infrastructure |
| H17 | A Rural Lens on a Research Agenda for Intelligent Infrastructure |
| J17 | Privacy in Information-Rich Intelligent Infrastructure |
| K17 | Smart Wireless Communication is the Cornerstone of Smart Infrastructures |

*This whitepaper is part of a series of papers compiled every four years by the CCC Council and members of the computing research community to inform policymakers, community members and the public on important research opportunities in areas of national priority. The topics chosen represent areas of pressing national need spanning various subdisciplines of the computing research field. The whitepaper attempts to portray a comprehensive picture of the computing research field detailing potential research directions, challenges and recommendations.*

*This material is based upon work supported by the National Science Foundation under Grant No. 1734706. Any opinions, findings, and conclusions or recommendations expressed in this material are those of the authors and do not necessarily reflect the views of the National Science Foundation.*